\documentclass[12pt]{article}

\usepackage{epsfig}

\usepackage{amssymb}
\usepackage{amsfonts}

\usepackage{color}
\def\be{\begin{equation}}
\def\ee{\end{equation}}
\def\ba{\begin{array}{c}}
\def\ea{\end{array}}

\newcommand{\bea}{\begin{eqnarray}}
\newcommand{\eea}{\end{eqnarray}}

\newcommand{\pbr}{\prec\!\!}
\newcommand{\pkt}{\!\!\succ\,\,}
\newcommand{\kt}{\rangle}
\newcommand{\br}{\langle}

\newcommand{\bbr}{\br\!\br}

\newcommand{\dkt}{]\!]}
\newcommand{\dbr}{[\![}

\begin{document}

\begin{center}

.

{\Large \bf

Non-stationary quantum mechanics
in hybrid non-Hermitian interaction representation

 }

\vspace{10mm}

\vspace{0.2cm}

\vspace{2mm}

\textbf{Miloslav Znojil}

\vspace{0.2cm}

\vspace{0.2cm}

Department of Physics, Faculty of Science, University of Hradec
Kr\'{a}lov\'{e},

Rokitansk\'{e}ho 62, 50003 Hradec Kr\'{a}lov\'{e},
 Czech Republic

\vspace{0.2cm}

 and

\vspace{0.2cm}

Institute of System Science, Durban University of Technology,
Durban, South Africa

 and

\vspace{0.2cm}

The Czech Academy of Sciences, Nuclear Physics Institute,

 Hlavn\'{\i} 130,
250 68 \v{R}e\v{z}, Czech Republic

\vspace{0.2cm}

{e-mail: znojil@ujf.cas.cz}

\end{center}

%\newpage

\section*{Abstract}

A
new formulation of unitary quantum mechanics
is proposed. Conceptually
our proposal is inspired
not only by the conventional
interaction picture
(where the wave functions
{\em and\,} the
observables do {\em both\,}
vary with time)
and by the newer,
non-Hermitian version of Schr\"{o}dinger picture
(where the Hamiltonians
with real spectra are still stationary)
but also by their universal non-Hermitian
interaction-picture ``overlap'' (NIP)
in which the physical
inner-product metric in Hilbert space
is {\em also\,} time-dependent.
The main idea of
the innovation
(called ``hybrid interaction picture'', HIP)
lies in a {\em partial\,}
simplification of the
NIP Hilbert-space metric
achieved via a {\em partial\,}
preconditioning of
the operators (in this sense we call their necessary
Hermitization ``hybrid'').
Via an illustrative example it is shown that the
HIP approach might open new ways towards the technically
optimal quantum-model-building strategies.

%\newpage

\section*{Keywords}
.

non-Hermitian quantum mechanics of closed systems;

non-stationary physical inner product in Hilbert space;

simultaneous simplification of the observables and of the metric;

solvable two-state model;

\newpage

\section{Introduction}

The economy of predictions of
the measurable quantitative properties of
bound states of
the so called
closed (i.e., unitary) quantum systems
often leads to a strongly preferred use of
the formulation of
quantum mechanics in the conventional textbook
Schr\"{o}dinger picture (SP, \cite{Messiah}).
In this approach
(with the quantum system
characterized by one or several self-adjoint and
time-independent
operators of observables)
the
evolution of a pure state $\psi(t)$
is fully determined by
the time-dependence of the representative ket-vector element
$|\psi(t)\pkt $
of a suitable physical SP Hilbert space ${\cal H}_{textbook}$
(for additional relevant details see section \ref{sekcedva} below).

In 1956, the latter, traditional model-building philosophy
appeared shattered by Freeman Dyson \cite{Dyson}.
Purely empirically he revealed that a non-standard, stationary but
non-unitary preconditioning
of a given conventional
but next-to-prohibitively complicated self-adjoint Hamiltonian
$\mathfrak{h}$ (yielding its simpler
but non-Hermitian isospectral
partner $H$)
may lead to a
perceivable simplification of
calculations (see a summary of
some of the related technicalities
in section~\ref{sekcetreti} below).

During the years of subsequent developments
the Dyson-inspired theory became fairly well
understood \cite{Geyer}. In
the stationary dynamical setup
(of what could be called a non-Hermitian Schr\"{o}dinger
picture, NSP) and in
a way described also in
newer reviews \cite{Carl,ali,book} the
various
manifestly non-Hermitian
quantum models have been shown compatible
with the traditional and
conventional quantum physics textbooks
as well as with the
related mathematical theorems~\cite{Stone}.

The
basic NSP idea can be found comparatively straightforward
because it is just
sufficient to
modify ``the inner product
of the Hilbert space properly'' \cite{ali}.
Unfortunately,
the practical applicability of the
non-Hermitian model-building recipes
is not too robust. Even in the simpler
stationary NSP scenario
the success appeared to
depend strongly on the properties of
the specific preselected sets of the observables.
One may notice that
the whole chapter
of review \cite{ali}
had to be devoted to
the techniques of calculation
of the necessary, auxiliary and nontrivial, ``Hermitizing''
inner-product metric $\Theta$.
Due to this technical obstacle,
physicists using the theoretical NSP framework often
restricted their attention just to the ``specific models
that allow for an explicit investigation'' (\cite{ali}, p. 1224).
This, fortunately, enabled them to
circumvent various immanent mathematical obstacles
(cf., e.g., their sample discussed in \cite{Dieudonne,SKbook}).

One of the
typical pragmatic, physics-oriented solutions of the latter
mathematical problem
can be found proposed
in \cite{Geyer} where
the authors restricted the
``mathematically admissible'' scope of the NSP theory to
the unitary quantum models with the bounded-operator
representations of the observables.
Under the important simultaneous
assumption of the time-independence of the
inner-product metric $\Theta$ we
also contributed,
in papers \cite{hybrid} and \cite{Nhybrid},
to these developments.
While still
not leaving
the most economical framework of the stationary NSP
formulation of quantum mechanics
we
proposed
there
the two non-standard
versions of the NSP approach.
For the reasons which will be explained below
we called them ``hybrid''.

On these grounds it was possible to
re-address several new classes of various realistic
physical problems and methods
(see, e.g., \cite{Bishop} for references).
The
systematic amendments of mathematics
opened the way towards treating
even the innovative NSP formalism itself as over-restrictive.
Successfully, attention has been redirected
towards the non-Hermitian Heisenberg picture (NHP,
in which the physical Hilbert-space metric was still
time-independent \cite{NHeisenberg},
i.e., not of our present interest)
and, ultimately, towards the most flexible
non-Hermitian interaction picture (NIP, \cite{NIP})
in which even
the physical inner product itself
is allowed to be non-stationary.

Not too surprisingly, the NIP language appeared to offer
not only the universal but also the most technically complicated
theoretical framework
(see section \ref{sekcectvrta} for its compact outline).
In it,
the
closed and unitary quantum systems are
characterized by the sets of
manifestly non-Hermitian and explicitly time-dependent
operators of observables
as well as by the particularly sophisticated,
manifestly
non-stationary Hermitian-conjugation metric
$\Theta(t)$.

The ``ultimate'' NIP
generalization of the formalism
appeared to lead to the emergence
of certain new conceptual as well as
purely technical obstacles.
In the present paper we intend to propose a next-step reformulation
of the NIP quantum mechanics, therefore.
The main theoretical aspects of our message will be
explained in section \ref{sekcepata}.
We will combine there the overall NIP background with an
incorporation of the above-mentioned ``hybrid construction''
philosophy.
The details will be explained below.
At the moment let us only
introduce an {\it ad hoc\,} notation convention
by which we will distinguish between the stationary and non-stationary
operators. The latter ones, Hermitian or non-Hermitian,
will be distinguished
by the presence of the argument $t$ (=time)
like, e.g., in the symbol
$\Theta(t)$ for the physical Hilbert-space
metric.
In contrast, the omission of the
symbol of time, say, in the Dyson-map factor
$\Omega$ of stationary metric
$\Theta=\Omega^\dagger\,\Omega$
will be supposed to imply
the time-independence of
$\Omega(t)=\Omega(0)$
as well as of
$\,\Theta(t)=\Theta(0)$, respectively.

Let us repeat that
the presentation of our results will
be preceded, in section \ref{sekcedva}, by
a concise reference
to the conventional
textbook formulations of quantum mechanics
in its Schr\"{o}dinger and Heisenberg representations
{\it alias\,} pictures.
After an equally concise recollection of the
two most common
non-Hermitian
but stationary versions of Schr\"{o}dinger picture in
section \ref{sekcetreti}
we will
turn attention to a recollection of
its formally equivalent non-Hermitian
interaction picture alternative (NIP) in section \ref{sekcectvrta}.
The detailed description of our new,
non-stationary and hybrid
reformulation
of quantum mechanics
will subsequently follow
in sections \ref{sekcepata} and \ref{sekcenova}.
We will argue that whenever
the standard NSP and/or NIP
methods of the description of quantum dynamics
are found prohibitively complicated, a remedy
may still be sought in a suitable factorization
of the
time-dependent Dyson-map operator ${\Omega}(t)$,
 \be
 {\Omega}(t)={\Omega}_2(t)\,{\Omega}_1(t)\,.
 \label{[19]}
 \ee
In the context of applications,
our present methodical proposal will be shown motivated by
the currently encountered technical difficulties
of description of the time evolution of the
closed and unitary quantum systems which are
characterized by certain sets of
manifestly non-Hermitian and explicitly time-dependent
operators of observables
(see, once more, review paper \cite{Bishop} for references).
Some of the specific merits of the resulting
hybrid NIP (HIP) model-building strategy
will finally be illustrated,
via a schematic model, in section \ref{sekcesesta}
and summarized in section \ref{sekcesedma}.

\section{Conventional formulations of
quantum mechanics\label{sekcedva}}

%\subsection{Schr\"{o}dinger picture}

In the most common Schr\"{o}dinger's formulation of quantum
mechanics (see, e.g., its concise outline in \cite{Styer})
the state of a closed and unitary quantum system ${\cal S}$ is
described by a time-dependent
ket-vector element $|\psi\pkt=|\psi(t)\pkt$ of a suitable
Hilbert space ${\cal H}_{textbook}$
\cite{Messiah}. This means that the description of the
evolution of the system
requires just the solution of
Schr\"{o}dinger equation
 \be
 {\rm i}\frac{d}{dt}\,|\psi(t)\pkt=H_S\,|\psi(t)\pkt\,,
 \ \ \ \ |\psi(t)\pkt \in {\cal H}_{textbook}\,.
 \label{tdse}
 \ee
All of the observable properties of ${\cal S}$ are
represented by the self-adjoint and, usually, stationary
operators
(denoted, say, as $A_S\neq A_S(t)$).
The
prediction of the results of the measurements is probabilistic,
based on the evaluation of the
time-dependent matrix elements
$\pbr \psi(t)|\,A_S\,|\psi(t) \pkt\,$.
For the guarantee of the unitarity
of the evolution as described by Eq.~(\ref{tdse})
the generator {\it alias\,}
Hamiltonian (which also represents the
energy
of the system)
must necessarily be, due to the
Stone theorem \cite{Stone}, self-adjoint in
${\cal H}_{textbook}$,
$H_S=H_S^\dagger$.

%\subsection{Heisenberg picture}

The best known alternative formulation of quantum
mechanics (carrying the Heisenberg's name, see the same
review \cite{Styer} for
the original references)
may be most easily obtained
via the following unitary-transformation ansatz
for the Schr\"{o}dinger's wave-function ket vector,
 \be
 |\psi(t)\pkt
 = {\mathfrak u}(t)\,|\psi_H(t)\pkt\,,
 \ \ \ \  {\mathfrak u}(t)=\exp (-{\rm i}\,(t-t_0)\,H_S)\,.
 \label{Hdefi}
  \ee
This ansatz leads to the three remarkable consequences.
First,
the above-mentioned Schr\"{o}dinger equation becomes
drastically simplified,
 \be
 {\rm i}\frac{d}{dt}\,|\psi_H(t)\pkt=0\,,
 \ \ \ \ |\psi_H(t)\pkt =|\psi_H(0)\pkt
 \in {\cal H}_{textbook}\,.
 \label{Htdse}
 \ee
Second, although
the generator of evolution of the new wave function
is often called Hamiltonian in the literature,
such a convention may be
misleading because in (\ref{Htdse}),
the so-defined ``false Hamiltonian'' would be
equal to zero identically.
Third,
Schr\"{o}dinger Eq.~(\ref{tdse}) appears redundant and
may be declared solved
immediately
after an arbitrary initial choice
of the
$H-$subscripted constant state-vector.

The change (\ref{Hdefi}) of the
representation of the ket vectors must be
accompanied by the simultaneous change of the representation
of all of the relevant observables,
 \be
 A_S \ \to \ \widetilde{A}_H(t) =  {\mathfrak u}^\dagger(t)\,
 A_S \ {\mathfrak u}(t)\,.
 \label{defop}
 \ee
The need of a careful terminology
is dictated by the ${\mathfrak u}(t)-$induced
time-dependence of $\widetilde{A}_H(t)$ in Eq.~(\ref{defop}).
An elementary check shows that
 \be
 {\rm i}\frac{d}{dt}\,\widetilde{A}_H(t)=
 \widetilde{A}_H(t)\,H_S
 -H_S\,\widetilde{A}_H(t)\,.
 \label{Heieq}
 \ee
Relation (\ref{Heieq})
is usually interpreted as
the Heisenberg evolution equation for observables.
One can conclude that we have
the correspondence
 \be
  \pbr \psi(t)|\,A_S\,|\phi(t) \pkt\,=\ \,
  \pbr \psi_H|\,\widetilde{A}_H(t)\,|\phi_H \pkt\,.
 \label{identi}
 \ee
As expected, all of the predictions of all of the
meaningful experimental measurements as calculated
in the respective Heisenberg and Schr\"{o}dinger picture
remain the same.
Moreover, due to the incidental coincidence  of
$H_S$ with its transform ${\mathfrak u}^\dagger(t)\,
H_S\,{\mathfrak u}(t)$,
the generator of evolution in (\ref{Heieq}) coincides with the
operator
representing the energy.
This is a rather lucky coincidence,
due to which no confusion is introduced when
people keep calling
the latter self-adjoint
operator a ``Hamiltonian in Heisenberg picture''.
A confusion can only emerge after a non-Hermitian
generalization of the conventional Heisenberg picture
(see \cite{NHeisenberg}).

\section{Non-Hermitian Hamiltonians in Schr\"{o}dinger picture\label{sekcetreti}}

Schr\"{o}dinger
equation (\ref{tdse})
describes
the time-evolution of a closed and unitary quantum system
if and only if the Hamiltonian operator
is self-adjoint in its Hilbert space
${\cal H}_{textbook}$ \cite{Stone}.
For this reason, it first sounded like a paradox
when Bender with Boettcher
proposed that
for certain unitary and closed quantum systems
the Hamiltonian
need not
necessarily be self-adjoint \cite{BB}.
The apparent puzzle has quickly been clarified
because
Bender with Boettcher
defined their Hamiltonian
in a manifestly unphysical
Hilbert space ${\cal H}_{auxiliary}$ \cite{ali}.
The subsequent resolution of the puzzle was
easy \cite{Carl}: One only had to replace ${\cal H}_{auxiliary}$
by another,
correct
Hilbert space ${\cal H}_{physical}$
which would be unitarily equivalent
to ${\cal H}_{textbook}$.
A vivid interest in the resulting non-Hermitian version of the
conventional Schr\"{o}dinger picture has been born.

The difference between the spaces  ${\cal H}_{auxiliary}$
and ${\cal H}_{physical}$
was only given
by the difference in the respective inner products.
At present, the details may already be found described
in multiple reviews (see, e.g.,
\cite{Geyer,Carl,ali,book,SIGMA}).
Currently, it is widely accepted that
in non-Hermitian
Schr\"{o}dinger picture
the evolution of a closed and unitary
quantum system ${\cal S}$ can be controlled
by a NSP Hamiltonian which
is non-Hermitian in ${\cal H}_{auxiliary}$.
At the same time, the NSP inner products themselves remain
stationary, time-independent.

From the historical perspective \cite{MZbook}
it is worth adding that
the manifestly non-Hermitian but still stationary
representation of a quantum Hamiltonian
has already been
used by Dyson \cite{Dyson}. He imagined that
a non-unitary change of the Hilbert space
could serve as
a promising source of simplifications.
In an
overall Schr\"{o}dinger-picture framework
he preconditioned the states using the ansatz
 \be
 |\psi(t)\pkt
 = \Omega\,|\psi(t)\kt\,\in \,{\cal H}_{textbook}
 \,,
 \ \ \ \ |\psi(t)\kt \in {\cal H}_{auxiliary}
 \,
 \label{stNHdefi}
  \ee
in which the mapping
${\cal H}_{textbook} \to {\cal H}_{auxiliary}$
is assumed stationary but non-unitary,
 \be
  \Omega \neq \Omega(t)\,,\ \ \ \
 \Omega_{} \neq
 \left ( \Omega_{}^{-1} \right)^\dagger\,.
 \label{stDmap}
 \ee
All of the operators of observables
were also assumed time-independent
(or, at worst, quasi-stationary - cf., e.g., \cite{Figuera} or
Theorem 2 in \cite{ali}).
This led to a comparatively
straightforward equivalent NSP reformulation of quantum mechanics in which
the new, hiddenly Hermitian Hamiltonian
 \be
 H = \Omega^{-1}\,H_S\,\Omega\ \neq\ H^\dagger
 \,
 \label{loss}
 \ee
still represented the observable energy.

As long as the transformation $\,H_S \to H\, $ is isospectral, it
enables one to replace the standard
Schr\"{o}dinger
Eq.~(\ref{tdse})
by its
equivalent form
 \be
 {\rm i}\frac{d}{dt}\,|\psi(t)\kt=H\,|\psi(t)\kt\,,
 \ \ \ \ |\psi(t)\kt \in {\cal H}_{auxiliary}\,,
 \ \ \ \ H
 \neq H^\dagger\,.
 \label{nhtdse}
 \ee
The new Hamiltonian $H$
characterized by its quasi-Hermiticity
\cite{Geyer,Dieudonne}
property
 \be
 H^\dagger\,\Theta=\Theta\,H\,,\ \ \ \ \
 \Theta=\Omega^\dagger\,\Omega\,,
 \label{quaqua}
 \ee
is often called, by physicists, ``non-Hermitian''.
A less misleading name of the property
would be, e.g., ``crypto-Hermiticity'',
mainly due to the coincidence of the
matrix element overlaps
 \be
  \pbr \psi(t)|\,H_S\,|\phi(t) \pkt
 =
 \br \psi(t)|\,\Theta\,H\,|\phi(t)\kt\,.
 \label{NNidenti}
 \ee
This is a identity which
could be re-read as a metric-based re-Hermitization
of~$H$.

One can conclude
that the
reinstallation of the conventional
physics is achieved by
the mere amendment of the inner product, i.e.,
via an amendment
${\cal H}_{auxiliary}\to{\cal H}_{physical}$ of the
Hilbert space.
Schematically, the underlying NSP
theoretical pattern
can be characterized by the following
``three-Hilbert-space'' diagram
as proposed in \cite{SIGMA},
 \be
 \label{u2hum}
  \ba
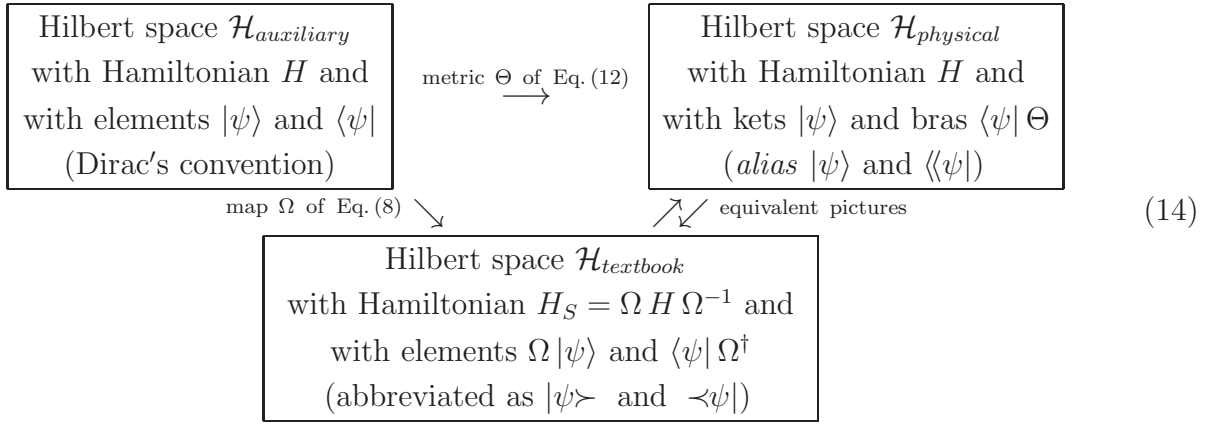
   \begin{array}{|c|}
 \hline
  {\rm Hilbert\  space\  }{\cal H}_{auxiliary} \ \\
 \ {\rm with\  Hamiltonian\ } H \ {\rm  and\ }
 \\
  {\rm  with\ }
 {\rm  elements\ } |\psi_{}\kt
  \ {\rm  and\ }
 \br \psi_{} | \\
   {\rm (Dirac's\ convention)  }\\
 \hline
 \ea
 \
 \
 \stackrel{{\rm metric}\ \Theta\ {\rm of\ Eq.\, (12)}}{ \longrightarrow }
 \
 \begin{array}{|c|}
 \hline
  {\rm  Hilbert\  space\    }{\cal H}_{physical} \ \\
 \ {\rm with\ Hamiltonian\ } H \ {\rm  and\ }
 \\
  {\rm  with\ }
 {\rm  kets\   } |\psi_{}\kt
  \ {\rm  and\ bras\ }
 \br \psi_{} |\, \Theta \\
 ({\it alias\   } |\psi_{}\kt
  \ {\rm  and\ }
 \bbr \psi_{} |)\, \\
 \hline
 \ea
 \
 \\ \ \ \ \ \ \
 \stackrel{{\rm map\ }\Omega_{}\ {\rm of\ Eq.\, (8)}}{ } \   \searrow \ \ \ \ \ \ \ \ \ \ \ \
 \ \ \ \ \ \ \ \  \nearrow \! \swarrow  \ \stackrel{{\rm equivalent\ pictures
 \ }}{ }
   \\
 \begin{array}{|c|}
 \hline
  {\rm  Hilbert\ space\   }{\cal H}_{textbook} \ \\
 \ {\rm with\ Hamiltonian\  } H_S=\Omega\,H\,\Omega^{-1} \ {\rm  and\ }
 \\
  {\rm  with\ }
 {\rm  elements\  }\Omega\, |\psi_{}\kt
  \ {\rm  and\ }
 \br \psi_{} |\,\Omega^\dagger \\
 ({\rm abbreviated\ as\    } |\psi_{}\pkt
   {\rm  and\ }
 \pbr \psi_{} |)\, \\
 \hline
 \ea
 \\
 \ea
 \\
 \ee
In such a diagramatic representation of the NSP theory
a direct return to the quantum mechanics of
textbooks is realized in the lowermost box.

Alternatively, a formally equivalent picture of physics
is provided by the rightmost box of the diagram.
The equivalence is
mediated by the isospectrality
$\,H_S \sim H\, $ yielding
the guarantee of the
(formally, hidden) Hermiticity of $H$. This
enables us to speak
not only about the conventional representation of
physics in the lowermost box
(one might call it an ``option-A Hermitization'')
but also.
about  an ``option-B Hermitization''
in the rightmost box of the diagram.
Using such an abbreviation of terminology
the NSP formalism can be summarized
using the compact
Table~\ref{osijex}.

\begin{table}[h]

\caption{Two alternative
physical interpretations
of the non-Hermitian
Schr\"{o}dinger picture. Constructively one can proceed either
via the reference to the manifestly self-adjoint $H_S$
(option ``A''), or via the construction of
a Hermitizing inner-product metric $\Theta$
(option ``B'').}
 \label{osijex} \vspace{.4cm}
\centering
\begin{tabular}{||l|ccccc||}
    \hline \hline
   & consistent &
  & initial && consistent\\
 & option A&
 %$\longleftarrow$
 &setup &
 %$\longrightarrow$
 & option B
 \\
 \hline
 %\hline
 inner product
 metric &$I$&& $I$&$\longrightarrow$& $\Theta=\Theta_H=\Omega^\dagger\,\Omega$
 \\
 %\hline
 Hamiltonian&
 $H_S
 =\Omega\,H\,\Omega^{-1}=H_S^\dagger$&$\longleftarrow$
 & $H\neq H^\dagger$&& $H=\Theta^{-1}\,H^\dagger\,\Theta:=H^\ddagger$
 \\
 %\hline
 Hilbert space&
 ${\cal H}_{textbook}$&& ${\cal H}_{auxiliary}$&&${\cal H}_{physical}$
 \\
\hline \hline
\end{tabular}
\end{table}

%\subsection{Two-step Dyson maps $\Omega=\Omega_2\,\Omega_1$
%and the third ``option C''}

The results of measurements
find their correct prediction and
probabilistic interpretation via option A
(i.e., via an ultimate return to ${\cal H}_{textbook}$)
or via option B
(based on a replacement of ${\cal H}_{textbook}$ by
${\cal H}_{physical}$).
In practice, the latter formulation of
quantum mechanics
was, incidentally, the mainstream choice
in the recent research \cite{Carl,ali}.
Besides its novelty it also appeared to
offer an optimal model-building strategy whenever
there appeared good reasons for
a Dyson-map-mediated
transition
from a conventional
but user-unfriendly Hilbert space ${\cal H}_{textbook}$
to another, mathematically
more appealing alternative ${\cal H}_{auxiliary}$.

Technically, the
stationarity assumption (\ref{stDmap})
was
considered, for a long time, unavoidable.
Once we accepted this philosophy we
decided to amend the formalism slightly \cite{EPJP,PLa,EPL}.
In particular, in \cite{hybrid} we
introduced another auxiliary Hilbert space
${\cal H}_{intermediate}$ and, on this basis, we
described
a new version of the ``stationary-metric''
NSP quantum mechanics
in its amended,
non-Hermitian but still entirely traditional
Schr\"{o}dinger representation.

With the further details to be found in {\it loc. cit.},
let us only add that the resulting ``hybrid''
form of the stationary NSP
(to be called ``option C'', with its basic
formal features summarized in Table \ref{rexijex})
can in fact be perceived as one of the main sources of
inspiration of our present paper.
Only on this ``option C'' {\it alias\,} ``hybrid''
conceptual background
we were ultimately able to decide, in what follows,
to abandon
the stationary-metric NSP
framework as over-restrictive.

\begin{table}[h]

\caption{Non-Hermitian
Schr\"{o}dinger picture
using hybrid
Hermitization (option ``C'').
Factorized stationary Dyson map $\Omega=\Omega_2\,\Omega_1$,
a reduced inner-product
metric
$\Theta_2=\Omega_2^\dagger\,\Omega_2$
and the amended Hamiltonian $H_1
 =\Omega_1\,H\,\Omega^{-1}_1$ acting in
 a new, {\it ad hoc\,} Hilbert space ${\cal K}_{physical}$
are employed \cite{hybrid}.}
 \label{rexijex} \vspace{.4cm}
\centering
\begin{tabular}{||l||ccccc||}
    \hline \hline
  two-step Hermitization & data &
  &  change of metric &&  change of operator \\
  %
% & setup&
% &metric &
% & Hamiltonian
% \\
 \hline
 \hline
 inner product
 metric&$I$&$\rightarrow$& $\Theta_2$ && $\Theta_2$
 \\
 \hline
 observable (Hamiltonian)&$H\neq H^\dagger$&&  $H\neq \Theta_2^{-1}\,H^\dagger\,\Theta_2$
 &$\rightarrow$&$H_1
 =\Theta_2^{-1}\,H_1^\dagger\,\Theta_2$
 \\\hline
 Hilbert space&
 ${\cal H}_{auxiliary}$&& ${\cal H}_{intermediate}$
 &&${\cal K}_{physical}$
 \\
\hline \hline
\end{tabular}
\end{table}

\section{Non-Hermitian interaction picture
\label{sekcectvrta}}

In the papers on
non-Hermitian reformulation of quantum mechanics
in which a
non-stationary generalization ${\Theta}(t)$
of the metric has been used
(cf. \cite{SIGMA,timedep}
and also \cite{NIP,Wang,IJTP,IJTPa,IJTPb,ju19})
it was necessary to
leave the Schr\"{o}dinger picture.
Even the
non-Hermitian version of Heisenberg picture
as proposed in \cite{NHeisenberg}
appeared insufficient.
For consistency reasons it
required the time-independence of the
inner-product metric (cf. also \cite{NHeisenbergb}).
The
work in the more complicated
non-Hermitian interaction picture (NIP) framework
proved unavoidable.

A guide to the NIP reformulation of quantum mechanics
can be found in
the Dyson's proposal of stationary mapping (\ref{stNHdefi}).
The concept had to be
generalized by allowing the map to be time-dependent,
 \be
 |\psi(t)\pkt
 = {\Omega}(t)\,|\psi(t)\kt\,,
 \ \ \ \ |\psi(t)\kt \in {\cal H}_{auxiliary}
 \,.
 \label{NHdefi}
  \ee
One of the main consequences of
such an ansatz is that
its insertion
in the conventional Schr\"{o}dinger
Eq.~(\ref{tdse}) yields a very special
evolution equation for
the ket vectors in ${\cal H}_{auxiliary}$ \cite{SIGMA},
 \be
 {\rm i}\frac{d}{dt}\,|\psi(t)\kt=
 {G}(t)\,|\psi(t)\kt\,,
 \ \ \ \ {G}(t)={H}(t)-{\Sigma}(t)\,,\ \ \ \ \
 {\Sigma}(t)=\frac{{\rm i}}{{\Omega}(t)}\frac{d}{dt}\,{\Omega}(t)
 \,.
 \label{nhHtdse}
 \ee
After the
transition to the
manifestly unphysical
Hilbert space ${\cal H}_{auxiliary}$ in (\ref{NHdefi}),
the observable Hamiltonian ${H}(t)$ itself does not
control the evolution of the ket-vectors anymore.
Such a task is transferred to the difference ${G}(t)$
between the instantaneous and ``Coriolis'' energy.
This makes the new Schr\"{o}dinger equation
different not only from its conventional Hermitian
version~(\ref{Htdse})
but also from
the stationary
non-Hermitian
Eq.~(\ref{nhtdse}).

The
generalization
is to be called non-Hermitian interaction picture, NIP.
It appeared
useful
in many-body coupled cluster calculations \cite{Bishop}
or
in relativistic quantum physics \cite{NIP}.
The non-stationary Dyson maps might even find a role
in quantum gravity \cite{NIPb,NIPc,NIPd,WDW}.
In
the mathematical framework of Riemannian geometry,
the time-dependent
Dyson maps ${\Omega}(t)$ were called there, for this reason,
the generalized vielbein operators \cite{ju22,juarX}.

In the NIP context the
operators of observables
 \be
  \widetilde{A}(t) =  {\Omega}^{-1}(t)\,
 A_S \,{\Omega}(t)\,
 \label{ndefop}
 \ee
are time-dependent in general. This enables us to derive
the correct evolution equation for
the observables in ${\cal H}_{auxiliary}$.
Under the ``initial Hermitian stationarity'' assumption
$A_S \neq A_S(t)$,
the goal is achieved when one
differentiates expression (\ref{ndefop}) with respect to time.
The resulting
operator-evolution equation
 \be
 {\rm i}\frac{d}{dt}\,\widetilde{A}(t)=
 \widetilde{A}(t)\,{\Sigma}(t)
 -{\Sigma}(t)\,\widetilde{A}(t)\,
 \label{nHeieq}
 \ee
(usually called Heisenberg equation) is
an analogue of its Hermitian-theory predecessor (\ref{Heieq}).
The characteristic change is that
the role of the generator
is {\em not\,} played now by any ``false Hamiltonian'' (i.e., by $H_S$ or
by ${H}(t)$)
or by the wave-function-evolution generator ${G}(t)$
but by another, different, ``Coriolis force'' operator ${\Sigma}(t)$
(cf. Table \ref{besijex}).

\begin{table}[h]

\caption{Connection between the Coriolis force and the non-stationary metric.
 %of Hamiltonian:
%Via the reconstruction of $H_S$
%(option ``A''), or via
%an amendment of the inner-product metric
%(option ``B'')..
}
 \label{besijex} \vspace{.4cm}
\centering
\begin{tabular}{||l||lll||}
    \hline \hline
   \multicolumn{1}{||c||}{concept} &\multicolumn{1}{c}{construction}
   &\multicolumn{1}{c}{decomposition}
   &\multicolumn{1}{c||}{occurrence}\\
    \hline

   Coriolis force &${\Sigma}(t)=
   {\rm i}{\Omega}^{-1}(t)\,\dot{{\Omega}}(t)$&${\Sigma}(t)={H}(t)-{G}(t)$&
 Eqs.~(\ref{nhHtdse}), (\ref{nHeieq})
   \\ metric ${\Theta}(t)$&${H}^\dagger(t)\,{\Theta}(t)={\Theta}(t)\,{H}(t)$
 &
 ${\Theta}(t)={\Omega}^\dagger(t)\,{\Omega}(t)$&
 Eq.~(\ref{nidenti})
 \\
 %inner product
% metric &unchanged, $I$&&trivial, $I$&&$\Theta=\Theta(H)\neq I$
% \\
\hline \hline
\end{tabular}
\end{table}

%%\newpage

%\subsection{The operator-evolution role of Coriolis force}

After the replacement of
Schr\"{o}dinger picture by interaction picture
the basic assumptions
of applicability of the theory remain analogous.
Firstly, the practical
solution of the updated versions of the Schr\"{o}dinger
and Heisenberg
equations has to be assumed
feasible. Secondly, the experimental
predictions
have to be deduced from the identity
 \be
  \pbr \psi(t)|\,A_S\,|\phi(t) \pkt
 =
 \br \psi(t)|\,{\Theta}(t)\,\widetilde{A}(t)\,|\phi(t)\kt\,.
 \label{nidenti}
 \ee
The parallels emerge with the Hermitian Eq.~(\ref{identi})
as well as with its non-Hermitian stationary successor
(\ref{NNidenti}).

%\section{HIP in applications}

%\subsection{Schr\"{o}dinger-equation paradox}

In a marginal remark let us add that
in our preceding text we used
the notion of a ``false Hamiltonian'' twice, viz,
in connection with the ``Schr\"{o}dingerian''
operator $G(t)$ in Eq.~(\ref{nhHtdse}) and
in connection with the
``Heisenbergian''
operator $\Sigma(t)$ in Eq.~(\ref{nHeieq}).
Naturally, in the NIP formalism one needs to work with
{\em both\,} of these operators
(none of which is observable) so that
it makes good sense to keep the name ``Hamiltonian''
reserved to their sum $\,H(t)=G(t)+\Sigma(t)\ $
(which is, by construction, observable).
In this sense, the
``single-Hilbert-space'' NSP unitary quantum mechanics
and its
``three-Hilbert-space'' \cite{SIGMA} NIP
generalization
remain {\em both\,} fully compatible with
the mathematical Stone theorem \cite{Stone}.

\section{Hybrid interaction picture\label{sekcepata}}

In Schr\"{o}dinger picture
both the metric an the Dyson map have to be,
for consistency reasons, time-independent.
Their non-stationary versions
${\Theta}(t)$ and ${\Omega}(t)$ can only occur, therefore,
in the broader and more flexible NIP framework.
This makes
the NIP-based constructions of
mathematically tractable models
attractive but also perceivably more
challenging.
In particular, the realistic applications
of the NIP approach
may prove difficult when we are not able
to perform the necessary
reconstruction of ${\Theta}(t)$
and/or
of ${\Omega}(t)$.
In the stationary case a ``hybrid NSP''
remedy has been proposed in \cite{hybrid}.
Now we intend to complement this proposal
by its extension to
quantum systems which are non-stationary.

In the conclusions of paper \cite{hybrid} we wrote
that the factorization
of the Dyson maps
might prove applicable and useful
also in non-stationary quantum models.
At the same time we considered the possibility purely
academical. Only later we revealed that
also the {\em non-stationary\,} factorization
ansatz (\ref{[19]})
might have its specific and important merits
in applications: Some of them will be
illustrated, after the
more detailed description of the HIP formalism,
in section \ref{sekcesesta} below.

\subsection{Notation conventions}

After one introduces
the time-dependent and factorized Dyson
operator of Eq.~(\ref{[19]}),
the message of Eq.~(\ref{NHdefi}),
i.e., the procedure of
the mapping of $ {\cal H}_{auxiliary}$ upon
${\cal H}_{textbook}\, $ can be perceived,
without approximations,
as an iterative, two-step process, with
 \be
 |\psi(t)\pkt={\Omega}_2(t)\,|\psi(t)]\,,
 \ \ \ \
 |\psi(t)]
 = {\Omega}_1(t)\,|\psi(t)\kt
  \,
 \label{rreNHdefi}
  \ee
where
$|\psi(t)\kt \in {\cal H}_{auxiliary}$ and
$|\psi(t)\pkt \in {\cal H}_{textbook}$.
Newly we introduced an ``intermediate'' ket-vector $|\psi(t)]$
belonging to some other, not yet specified Hilbert space
(temporarily, let us denote it by a tilded symbol $\widetilde{\cal K}$).

In a preparatory step of our considerations
let us assume,
tentatively, that
the first-step change
in (\ref{rreNHdefi}) is small,
 \be
 {\Omega}_1(t)\approx I\,.
 \label{huja}
 \ee
The full metric
${\Theta}(t)$
would then be well approximated by
its reduced form
 \be
 {\Theta}_2(t)={\Omega}_2^\dagger(t)\,{\Omega}_2(t)\,.
 \label{inpr}
 \ee
The explicit use of the latter
inner-product metric would, strictly speaking
(i.e., without approximations)
convert the initial, mathematically maximally
friendly Hilbert space ${\cal H}_{auxiliary}$
into another, perceivably less user-friendly
Hilbert space ${\cal H}_{intermediate}$
which would already become ``approximately physical''
(see the more detailed motivation of
introduction of such a space in \cite{PLa}).
In other words, in the light of
our temporary assumption (\ref{huja}),
the ultimate transition
to the consistent quantum mechanics
(in ${\cal H}_{physical}$)
could be then achieved by perturbation techniques.

In the second preparatory step
let us now imagine that the correct probabilistic
interpretation of the quantum system in question
is achieved not only in
${\cal H}_{physical}$
(where, as we explained above,
the construction of the correct metric $\Theta(t)$
guarantees the  observable-energy status of
the Hamiltonian $H(t)$)
but also in
${\cal H}_{textbook}$ where the
correct inner-product metric is
trivial,
equal to the
identity operator. This implies the very standard
antilinear
ket - bra correspondence in
${\cal H}_{textbook}$,
 \be
 {\cal T}_{textbook}:|\psi(t)\pkt \to \ \pbr \psi(t)| \,.
 \label{ttex}
 \ee
An analogous
triviality of the Hermitian-conjugation
ket - bra correspondence
is also encountered
in the other two
manifestly unphysical Hilbert spaces
${\cal H}_{auxiliary}$ and $\widetilde{\cal K}$,
 \be
 {\cal T}_{auxiliary}:|\psi(t)\kt \to \, \br \psi(t)|\,,
 \ \ \ \
 \widetilde{\cal T}_{}:|\psi(t)] \to \, [\psi(t)|
 \,.
 \label{taux}
 \ee
A more sophisticated relation between the
inner product and Hermitian conjugation
characterizes the other two Hilbert spaces
${\cal H}_{intermediate}$ and
${\cal H}_{physical}$.
The respective ket $\to$ bra
Hermitian-conjugation correspondences
are
controlled there by the respective
nontrivial operators of metric,
 \be
 {\cal T}_{intermediate}
 :|\psi(t)\kt \to
 \br \psi(t)|\,
 {\Theta}_2(t)
 \,,
 \ \ \ \
 {\cal T}_{physical}:|\psi(t)\kt \to  \br \psi(t)| \,
 {\Theta}(t) :=\bbr \psi(t)|\,.
 \label{tintphy}
 \ee
Naturally,
one could hardly work with such a multitude of Hermitian
conjugations. For this reason,
all of the
different results of these different conjugations
will be reinterpreted here as the mere
definitions
of the respective new, differently
denoted bra-vectors in the single representation
space
${\cal H}_{auxiliary}$.

Such a reduction
of the universality of notation
will prove efficient and useful since
in the
unique representation space
${\cal H}_{auxiliary}$ we will
be allowed to work just with
the single
antilinear operator ${\cal T}_{auxiliary}:={\cal T}$
of Eq.~(\ref{taux}).
This will enable us to treat, e.g., Eq.~(\ref{NHdefi})
as a mere definition of $|\psi(t)\pkt$
in ${\cal H}_{auxiliary}$. Similarly, the second item in
Eq.~(\ref{rreNHdefi}) will be
just the definition of another ket-vector $|\psi(t)]$
in ${\cal H}_{auxiliary}$.

An analogous reduction of the notation
will be
extended to the Hermitian conjugation of any operator ${\cal O}$
marked, in
${\cal H}_{auxiliary}$, just by the conventional superscripted dagger,
${\cal O}\to {\cal O}^\dagger$.

\subsection{The six-Hilbert-space lattice}

The main technical weakness of our preceding considerations
is that besides the second item in Eq.~(\ref{taux}),
one could also be interested in
its alternative version using metric (\ref{inpr}).
The ambiguity is nontrivial. Fortunately, a guide to its
fully consistent resolution may be again sought
in
the papers where the
Dyson mappings were stationary.

For this purpose
we only have to modify
the tilded Hilbert space $\widetilde{\cal K}$ properly.
Using
the terminology and notation
of
paper
\cite{hybrid}
and
the guidance by
paper \cite{Nhybrid}
we just define
a ``missing'' operator product
$\Omega_{21}(t)=\Omega_2(t)\,\Omega_1(t)\,\Omega_2^{-1}(t)$
and arrive at
a consistent scheme.
The result
may be given the form of
the following new,
``hybrid interaction picture''-presenting (HIP)
non-stationary-metrics descendant of
the stationary-metrics NSP
diagram (\ref{u2hum}),
 \be
 \label{zu2hum}
  \ba
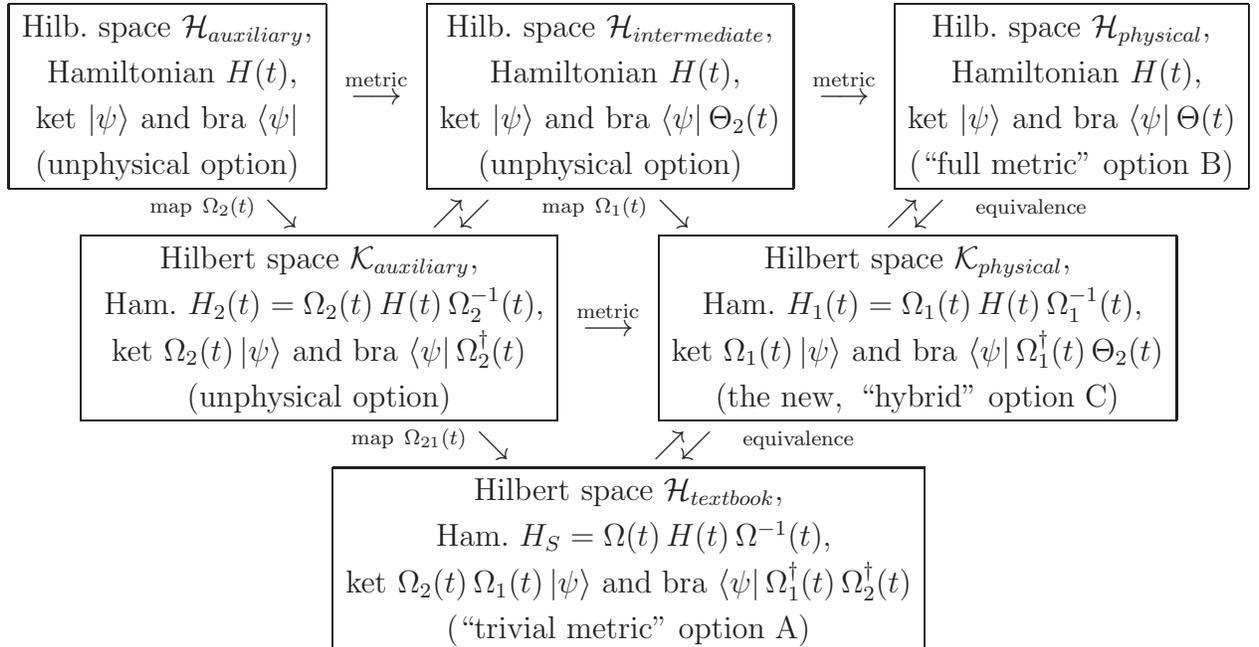
   \begin{array}{|c|}
 \hline
  {\rm  Hilb.\ space\    }{\cal H}_{auxiliary},\\
 \ {\rm  Hamiltonian\  } H(t),
 \\
 {\rm ket\ } |\psi_{}\kt
  \ {\rm  and\ bra\ }
 \br \psi_{} | \\
   {\rm  (unphysical\ option)  }\\
 \hline
 \ea
 \
 \stackrel{{\rm metric}}{ \longrightarrow }
 \
 \begin{array}{|c|}
 \hline
  {\rm  Hilb.\ space\    }{\cal H}_{intermediate},\\
 \ {\rm Hamiltonian\    } H(t),
 \\
 {\rm ket\  } |\psi_{}\kt
  \ {\rm  and\ bra\ }
 \br \psi_{} |\, \Theta_2(t) \\
  {\rm  (unphysical\ option)  }\\
 \hline
 \ea
 \
 \stackrel{{\rm metric}}{ \longrightarrow }
 \
 \begin{array}{|c|}
 \hline
  {\rm  Hilb.\ space\    }{\cal H}_{physical},\\
 \ {\rm  Hamiltonian\   } H(t),
 \\
 {\rm  ket\ } |\psi_{}\kt
  \ {\rm  and\ bra\ }
 \br \psi_{} |\, \Theta(t) \\
 {\rm (``full\ metric" \ option\ B)}\\
 \hline
 \ea
 \\
 \
 \stackrel{{\rm map\ }\Omega_2(t)}{ } \   \searrow \ \  \ \
 \ \ \ \ \ \ \ \ \   \nearrow \! \swarrow  \ 
 \stackrel{{\rm }}{ }
 \ \ \
  \
 \stackrel{{\rm  map\ }\Omega_1(t)}{ } \   \searrow \ \ \ \ \ \ \ \
 \ \ \ \ \  \ \ \ \ \ \ \nearrow \! \swarrow  \ \ \ \stackrel{{\rm equivalence
 \ }}{ } \ \ 
 \\
 \begin{array}{|c|}
 \hline
  {\rm  Hilbert\ space\    }{\cal K}_{auxiliary},\\
 \ {\rm   Ham.\ } H_2(t)=\Omega_2(t)\,H(t)\,\Omega_2^{-1}(t),
 \\
 {\rm ket\  }\Omega_2(t)\,  |\psi_{}\kt
  \ {\rm  and\ bra\ }
 \br \psi_{} |\,\Omega_2^\dagger(t) \\
  {\rm (unphysical\ option) }\\
 \hline
 \ea
 \
 \stackrel{{\rm metric}}{ \longrightarrow }
 \
 \begin{array}{|c|}
 \hline
  {\rm Hilbert\ space\     }{\cal K}_{physical},\\
 \ {\rm   Ham.\  } H_1(t)=\Omega_1(t)\,H(t)\,\Omega_1^{-1}(t),
 \\
 {\rm ket\  }\Omega_1(t)\, |\psi_{}\kt
  \ {\rm  and\ bra\ }
 \br \psi_{} |\,\Omega_1^\dagger(t)\, \Theta_2(t) \\
 {\rm (the \ new,
 \ ``hybrid"\ option\ C)}\\
 \hline
 \ea
 \\
 \stackrel{{\rm map\ }\Omega_{21}(t)}{ } \   \searrow \ \ \ \ \ \
 \ \ \ \ \ \ \ \  \nearrow \! \swarrow  \ \  \ \stackrel{{\rm equivalence
 \ }}{ }\ \ \ \
   \\
 \begin{array}{|c|}
 \hline
  {\rm Hilbert\ space\   }{\cal H}_{textbook},\\
 \ {\rm  Ham.\  } H_S=\Omega(t)\,H(t)\,\Omega^{-1}(t),
 \\
 {\rm ket\ }\Omega_2(t)\,  \Omega_1(t)\, |\psi_{}\kt
  \ {\rm  and\ bra\ }
 \br \psi_{} |\,\Omega_1^\dagger(t)\, \Omega_2^\dagger(t) \\
 {\rm (``trivial\ metric"\ option\ A)}\\
 \hline
 \ea
 \\
\ea
 \ee
The three rightmost boxes stand for the
three eligible consistent theoretical options representing
the unitary and probabilistic non-stationary quantum-model physics.
For this reason we will also use suitable abbreviations for the kets
 $ \Omega_2(t)\,
 \Omega_1(t)\, |\psi_{}\kt:=|\psi_{}(t)\pkt$ (cf. Eq.~(\ref{rreNHdefi}))
  and for the bras $ \br \psi_{} |\,\Omega_1^\dagger(t)\, \Omega_2^\dagger(t)
  :=\pbr \psi_{}(t) |$  (cf. Eq.~(\ref{ttex}))
of option A, for the
 bras $ \br \psi_{} |\, \Theta(t):=\bbr \psi_{}(t) |$
of option B, and for the kets
 $\Omega_1(t)\, |\psi_{}\kt:=|\psi_{}(t)]$ and for the bras
 $ \br \psi_{} |\,\Omega_1^\dagger(t)\, \Theta_2(t):= \dbr \psi_{}(t) |$
of option C.
These abbreviations were only introduced for
the rightmost triplet of
Hilbert spaces since only these spaces
are ``relevant'' and provide
the correct (i.e., eligible, equivalent) physical
representations of the quantum system in question.

In the six-space diagram (\ref{zu2hum}) the
key innovation lies in the middle line.
The inspection of its
two boxes reflects the fact that
our initial tentative introduction
of the tilded symbol $\widetilde{\cal K}$
was ambiguous. According to
Eq.~(\ref{rreNHdefi}) one
would expect the identification $\widetilde{\cal K}
={\cal K}_{physical}$. At the same time,
the second item in Eq.~(\ref{taux}) would only be compatible
with another, different
identification of $\widetilde{\cal K}
={\cal K}_{auxiliary}$.
Thus,  in place of
the single tilded letter
$\widetilde{\cal K}$,
two different
Hilbert-space symbols had to be introduced.

This is our final observation. It
enables us to conclude that
a consistent formulation of
quantum mechanics in its HIP representation
requires the reference
to as many as six different Hilbert spaces.
For the stationary NSP,
after all,
we arrived,
in \cite{hybrid} and \cite{Nhybrid},
not too surprisingly, at precisely
the same conclusion.
The main differences and innovation will
now emerge in connection with
the description of the evolution equations.

%\newpage

\section{The HIP evolution equations\label{sekcenova}}

From a less formal perspective
the inspection of diagram (\ref{zu2hum}) reveals
that the box with space ${\cal K}_{auxiliary}$
can be omitted as, in some sense, superfluous.
Secondly, once we decided to represent
all of the vectors and operators in
the mathematically most friendly Hilbert space
${\cal H}_{auxiliary}$, also the use of the box with
space ${\cal H}_{intermediate}$
can be skipped as purely formal.
Thus, for the practical, quantum-system-representation purposes
we are left just with the three rightmost,
``correct and physical'' indispensable
boxes.
Among them, precisely the middle one
containing Hilbert space ${\cal K}_{physical}$
and
Hamiltonian $H_1(t)$
is needed in
the HIP approach.
Indeed, only in this box
one makes use
of the factorized Dyson map (\ref{[19]}).

Let us now turn attention to the
new ket-vector elements $ |\psi(t)]={\Omega}_1(t)\,|\psi(t)\kt$
representing the states in the new Hilbert space
${\cal K}_{physical}$.
After the insertion
of ansatz (\ref{rreNHdefi}) in
(\ref{tdse})
we obtain the updated, HIP Schr\"{o}dinger equation
 \be
 {\rm i}\frac{d}{dt}\,|\psi(t)]=
 {G}_1(t)\,|\psi(t)]\,,
 \ \ \ \ {G}_1(t)={H}_1(t)-{\Sigma}_2(t)\,,\ \ \ \ \
 {\Sigma}_2(t)=\frac{{\rm i}}{{\Omega}_2(t)}\frac{d}{dt}\,{\Omega}_2(t)
 \,.
 \label{renhHtdse}
 \ee
It controls the evolution of the state-vectors.
In a close parallel to Eq.(\ref{nhHtdse}),
also the new, isospectral version
${H}_1(t)={\Omega}_1(t)\,{H}(t)\,{\Omega}_1^{-1}(t)$
of the energy-representing Hamiltonian
will not generate the evolution of the wave functions $|\psi(t)]$
just by itself.
Still, in comparison, the
unexpected merit of the
new equation is that it offers
a manifest separation of the information about energy
(carried now by our new Hamiltonian ${H}_1(t)$)
from the information about the geometry
carried by the partial Coriolis force ${\Sigma}_2(t)$,
by the dysonian ${\Omega}_2(t)$ and by
the metric (\ref{inpr}).

In the light of diagram (\ref{zu2hum})
the
new HIP Hilbert space
${\cal K}_{physical}$ shares the
nontrivial reduced metric (\ref{inpr})
with its
user-friendlier partner
${\cal H}_{intermediate}$.
The correct HIP description of dynamics is achieved
by the parallel
$\Omega_1(t)-$mediated amendment
$H(t) \to H_1(t)$ of the Hamiltonian.
For this reason, the new non-stationary HIP
formalism
based on the factorization (\ref{[19]}) and on the updated
Schr\"{o}dinger equation (\ref{renhHtdse})
deserves to be called ``hybrid''.

In contrast to the structure of the stationary NSP formalism
of Ref.~\cite{hybrid},
the new and inseparable part
of the HIP approach
becomes now the necessity of
the description of the time-dependence of the
observables.
In order to make such a feature of the formalism explicit,
we just have to insert the
factorized Dyson map in Eq.~(\ref{ndefop})
which defines the
time-dependent observable $\widetilde{A}(t)$
acting in ${\cal H}_{auxiliary}$ in terms of
its partner in Schr\"{o}dinger picture.
In
the HIP Hilbert space
${\cal K}_{physical}$ we may define operator
$\widetilde{A}_1(t)={\Omega}_1(t)
\,\widetilde{A}(t)\,{\Omega}_1^{-1}(t)$
as
an analogue
of Hamiltonian ${H}_1(t)$.
The standard NIP connection ${\Omega}(t)\, \widetilde{A}(t) =
A_S \,{\Omega}(t)$
between the observable $\widetilde{A}(t)$
and
its time-independent avatar $A_S$
can be now rewritten
as the specific HIP quasi-Hermiticity rule
 \be
 {\Omega}_2(t)\, \widetilde{A}_1(t) =
 A_S \,{\Omega}_2(t)\,.
 \label{rendefop}
 \ee
Both sides of this equation
can now easily be differentiated
with respect to time.
This yields relation
 \be
 {\rm i}\frac{d}{dt}\,\widetilde{A}_1(t)=
 \widetilde{A}_1(t)\,{\Sigma}_2(t)
 -{\Sigma}_2(t)\,\widetilde{A}_1(t)\,
 \label{renHeieq}
 \ee
i.e., the Heisenberg-type evolution equation
for the observable $\widetilde{A}_1(t)$.

The latter operator is,
due to relation (\ref{rendefop}),
quasi-Hermitian,
i.e., such that
 \be
 \widetilde{A}_1^\dagger(t)\,{\Theta}_2(t)={\Theta}_2(t)\,
 \widetilde{A}_1(t)\,.
 \ee
We arrive at the new, HIP version of the matrix-element
equivalence (\ref{nidenti}) which reads
 \be
  \pbr \psi(t)|\,A_S\,|\phi(t) \pkt
 =
 [ \psi(t)|\,{\Theta}_2(t)\,\widetilde{A}_1(t)\,|\phi(t)]
 =\dbr \psi(t)|\,\widetilde{A}_1(t)\,|\phi(t)]
 \,.
 \label{renidenti}
 \ee
Due to this identity the predictions
of the measurements evaluated in the NIP and HIP frameworks
are equivalent and indistinguishable.

The display of the two
alternative versions
of the right-hand-side overlap
in Eq.~(\ref{renidenti})
is not just a formal exercise because
in the light of comments as made in \cite{NIP}
the latter version is preferable and
mathematically optimal.
Indeed,
the necessity of the knowledge of the metric
${\Theta}_2(t)$ (i.e., of the time-dependent {\em operator})
is made redundant. In the formula its occurrence
merely specifies the
{\em vector\,} $|\psi(t)\dkt$, i.e., an
element of the dual
vector space ${\cal H}'_{intermediate}$
(marked by the prime).
This opens the way towards a reduction of the
calculations.
Their economy becomes enhanced
due to the possibility of a replacement
of the Heisenberg-like equation
for the metric ${\Theta}_2(t)$ by the
mere new Schr\"{o}dinger-like equation
for the ket-vector $|\psi(t)\dkt$.

The derivation of the latter, substitute  equation is straightforward.
We merely differentiate the
definition
of the
linear functional $\dbr \psi(t)|$
in the dual vector
space in ${\cal H}_{intermediate}'$
and arrive at the final ``second'' Schr\"{o}dinger equation
 \be
 {\rm i}\frac{d}{dt}\,|\psi(t)\dkt=
 {G}_1^\dagger(t)\,|\psi(t)\dkt\,.
 \label{ddtdse}
 \ee
Obviously, there is no need of constructing the
generator ${G}_1^\dagger(t)$
because its Hermitian conjugate is already known from the
``first''
Schr\"{o}dinger Eq.~(\ref{renhHtdse}).

%\subsection{Non-stationary two-state model}

\section{Non-stationary two-state toy model\label{sekcesesta}}

A clarification of the
practical aspects of the
above-outlined terminological
paradoxes
will be provided here via an
NIP- and HIP-based elementary example.
Our illustration of the
applicability of the formalism was inspired by
Ref.~(\cite{hybrid})
where we introduced
the
two-parametric factorizable metric operator
 \be
  \Theta_{}=
 %\left[ \begin {array}{cc} 1+{s}^{2}& \left( 1+{s}^{2} \right) r+s\\
% \noalign{\medskip}r+ \left( rs+1 \right) s& \left( r+ \left( rs+1
% \right) s \right) r+rs+1\end {array} \right]\,.
 \left[ \begin {array}{cc} 1+{s}^{2}&r+r{s}^{2}+s
 \\\noalign{\medskip}r+r{s}^{2}+s&{r}^{2}
 +{r}^{2}{s}^{2}+2\,rs+1\end {array} \right]\,
   \label{[219]}
 \ee
and where we proved (see Lemma Nr. 1 in {\it loc. cit.})
that this matrix is
positive definite. Via Eq.~(\ref{quaqua}) this
enabled us to construct a non-Hermitian but quasi-Hermitian
stationary Hamiltonian
matrix $H$ acting in
${\cal H}_{auxiliary}$
as well as its isospectral partner
$H_S$ which was Hermitian in
${\cal H}_{textbook}$.

For our present purposes we will use the same
metric (\ref{[219]})
but a slightly different, time-dependent Hamiltonian
matrices $H_S(t)$ and ${H}(t)$
acting in ${\cal H}_{textbook}$ and
${\cal H}_{auxiliary}$, respectively.
Both of them will
share the two real bound-state
energies,
viz.,
$E_0=E_0(t)=1+t$ and $E_1=2$,
only one of which is now kept
time-independent.
Only the former matrix
(viz., $H_S(t)$,
not of our present immediate interest) remains Hermitian.
In contrast,
the latter matrix, viz.,
 \be
 {H}(t)=
 \left[ \begin {array}{cc} -rs+rst+1+t&-{r}^{2}s+{r}^{2}st-r+rt
 \\ \noalign{\medskip}s-st&rs-rst+2\end {array} \right]
\label{ohjoj}
 \ee
is now chosen as a new, manifestly time-dependent solution
of the necessary quasi-Hermiticity constraint (\ref{quaqua})
with the old preselected metric (\ref{[219]}).

The parameters of
the model may vary with time.
First of all, the non-stationarity
will be enhanced
by the replacement of the constant parameter $s$
by the following elementary function of time,
 \be
 s=s(t)=a\,t+b\,t^2/2\,.
 \label{[29]}
 \ee
In the NIP formulation of quantum mechanics
this immediately leads to the purely imaginary
Coriolis-force matrix of Eqs.~(\ref{nhHtdse})
and (\ref{nHeieq}),
 \be
 {\Sigma}(t)=\left[ \begin {array}{cc}
  -ir \left( a+bt \right) &-ir \left( a+bt \right)
   \\\noalign{\medskip}i \left( a+bt \right) &i
    \left( a+bt
 \right) \end {array} \right]\,.
 \label{[30]}
 \ee
The knowledge of the two matrices ${H}(t)$ and ${\Sigma}(t)$
enables us to construct the ``false Hamiltonian''
generator ${G}(t)={H}(t)-{\Sigma}(t)$.
Although the latter matrix itself would look rather clumsy,
the computer-assisted
evaluation of its two eigenvalues remains straightforward
yielding the doublet
 \be
 \{1+t,2-ibt+ibtr-ia+iar\}.
 \label{[31]}
 \ee
Although the
underlying quantum system
is, by construction, unitary, we see that
the eigenvalues of the  ``false Hamiltonian''
${G}(t)$ are {\em never\,} real
(so that ${G}(t)$ {\em cannot\,} be quasi-Hermitian)
unless  $r=1$ or $r \neq 1$
and $a=b=0$
or $t=t_0\neq 0$
such that $bt_0+a=0$.
One can conclude that
there is no reason for calling the operator
${G}(t)$ a Hamiltonian.
The terminological
Schr\"{o}dinger-equation Hamiltonian-ambiguity
paradox is resolved.

In the methodological setting
it is important that our toy-model time-dependence
ansatz (\ref{[29]}) really
makes the metric (\ref{[219]})
manifestly time-dependent.
A less pleasant
byproduct of this ansatz is that
the insertion of formula (\ref{[29]})
would already make the explicit matrix forms
of the metric
as well as of the related Hamiltonian (\ref{ohjoj})
too long for display in print.
The
latter observation means that the
questions of
optimality of the choice
and of the factorization of Dyson map ${\Omega}(t)$
arise with new urgency.

In a methodical context the latter remark offers another
rather persuasive
argument in favor of transition
to the HIP formalism.
In the HIP context, in our present illustrative example,
{\em both\,} of the upgraded and most
relevant matrices ${\Theta}_2(t)$ and ${H}_1$
happen to be made {\em simultaneously\,}
simplified and printable, with
 \be
 {\Theta}_2(t)=
 \left[ \begin {array}{cc} 1+{s}^{2}(t)&s(t)
 \\\noalign{\medskip}s(t)&1\end {array} \right]
  \label{Z21}
 \ee
and
 \be
 {H}_1=\left[ \begin {array}{cc} 1+t&0
 \\\noalign{\medskip}s(t)-t\,s(t)&2\end {array} \right]\,.
 \ee
Obviously, although the latter Hamiltonian matrix is
non-Hermitian, its triangular form makes the
verification of its isospectrality with ${H}(t)$ trivial.
Secondly, the equally essential
proof of the positivity of the partial metric
(\ref{Z21}) is much easier than
the analogous, not too trivial
proof as provided,
for the full metric (\ref{[219]}),
in \cite{hybrid}.

Another argument supporting
transition from
the more conventional NIP approach to
its present HIP amendment concerns the
full Coriolis force ${\Sigma}(t)$ as
displayed in Eq.~(\ref{[30]}) above.
Indeed, in the NIP framework
the subtraction of this matrix
from the Hamiltonian ${H}(t)$
yielded the ``false Hamiltonian'' matrix ${G}(t)$
which
appeared to be too complicated to be displayed here
in print.

In the HIP framework
an analogous generator ${G}_1(t)$ of the evolution
of wave functions
is needed, first of all, in the HIP Schr\"{o}dinger-type
Eq.~(\ref{renhHtdse}). In fact, it was
truly impressive
to discover that
the latter, ``false Hamiltonian'' operator
remains to be, unexpectedly, next to elementary,
 \be
 {G}_1(t)=\left[ \begin {array}{cc} 1+t&0
 \\\noalign{\medskip}at-a{t}^{2}+1/2\,b{t}^{2}-1/2\,b{t}^{3}
 -i \left( a+bt \right) &2\end {array} \right]\,.
 \label{[41]}
 \ee
In a sharp contrast to the NIP scenario,
the eigenvalues of such an
``intermediate false Hamiltonian''
(which are equal to its diagonal matrix elements)
remain real.
Thus, the time-evolution of the new, HIP state vectors $|\psi(t)]$
is definitely not as wild as
it was in the old NIP scenario.

The evaluation of the ultimate
probabilistic HIP predictions
using formula (\ref{renidenti})
would also require the choice of the observable
of interest,
i.e., the specification
of its non-Hermitian-operator
representation $\widetilde{A}_1(t)$.
From its initial operator value $\widetilde{A}_1(t_0)$
postulated
at the time $t=t_0$ of the preparation of the system,
its
evaluation at $t>t_0$ must be performed
via the
solution of the HIP Heisenberg-type Eq.~(\ref{renHeieq}).
Incidentally, the
corresponding HIP Coriolis force
{\it alias\,} Heisenberg Hamiltonian
 \be
 {\Sigma}_2(t)=\left[ \begin {array}{cc} 0&0
 \\\noalign{\medskip}i \left( a+bt \right) &0\end {array} \right]
 \label{[42]}
 \ee
is certainly persuasively simpler than
its general NIP predecessor ${\Sigma}(t)$
of Eq.~(\ref{[30]}).

%\newpage

\section{Discussion and summary\label{sekcesedma}}

Our present HIP
reformulation of non-stationary
quantum mechanics can be perceived as
a certain climax of the lasting
efforts devoted to the search for
an optimal formulation of unitary quantum mechanics
using the operators of observables in
their mathematically
less restricted and more flexible
non-Hermitian
representations \cite{Geyer,Carl}.
During this search it has been revealed that
the technical aspects of such a generalization of the formalism
are important. Their fairly surprising
merits (offering multiple new insights in quantum phenomenology
\cite{Christodoulides,Carlbook})
appeared accompanied, indeed, by some not always expected
mathematical obstacles \cite{Trefethen,Siegl,Viola}.

In our paper we addressed the most general conceptual scenario
in which the scope of the quantum theory is, in some sense, maximal.
Admitting, in the Hilbert space of states,
the variability of the physical
inner product (i.e., of its operator metric $\Theta(t)$)
with time.
In essence, the proposed HIP approach
can be perceived as an amended version of the
NIP formulation of quantum mechanics \cite{NIP}.
In the innovation
the standard model-building strategy
(based on a brute-force reconstruction of $\Theta(t)$)
is made more flexible.
The flexibility (achieved via the freedom in factorization (\ref{[19]}))
is shown to imply, first of all, a partial separation
of the kinematics (represented by the partial metric $\Theta_2(t)$)
from the dynamics (represented,
typically, by the partially preconditioned observable $H_1(t)$).

One of the key mathematical merits of the present
non-stationary HIP amendment
of the theory can be seen in the close parallels with
its stationary hybrid NSP predecessor (cf. its
summary in Table \ref{rexijex})
as well as with the plain NIP approach
(cf. its outline in section \ref{sekcectvrta}).
In comparison, in a way well illustrated in section \ref{sekcesesta},
one may really expect
a perceivable simplification of the evolution equations
achieved after the change of framework from NIP to HIP:
{\it Pars pro toto\,} let us just recall the
drastic contrast between the persuasive user-friendliness
of the two HIP toy-model generators $G_1(t)$ (of Eq.~(\ref{[41]}))
and/or $\Sigma_2(t)$ (of Eq.~(\ref{[42]}))
and the enormous user-unfriendliness
of their respective NIP partners $G(t)$ (found to be
too large for the display in print)
and/or $\Sigma(t)$ of Eq.~(\ref{[30]}).

Besides the simplification of the structure
of the matrices, an equally important advantage of the
constructive HIP recipe is indicated by the
differences between the spectra of our
toy-model NIP and HIP generators.
Indeed, in contrast to the purely real spectra of
$G_1(t)$
and $\Sigma_2(t)$,
the analogous spectra of their respective
NIP partners $G(t)$
and $\Sigma(t)$ are both almost always
complex (cf. Eq.~(\ref{[31]})).
In practice, naturally, the imaginary
components of these spectra will
be directly responsible for the
emergence of
certain quickly (i.e., exponentially quickly)
growing
components of the
related (i.e.,
Schr\"{o}dinger and/or Heisenberg)
evolution-equation solutions.
In principle, it is true that the latter components will have to
cancel in the ultimate predictions (\ref{renidenti})
of the measurements,
but the numerical precision of the evaluation of these matrix
elements might be decisively influenced (i.e., damaged)
by the occurrence
of these (presumably, very large) exponentials.

%\newpage

\section*{Acknowledgements}

The present research project has been inspired by
the questions asked by an anonymous referee
of Ref.~\cite{hybrid}.

\end{document}